**Investigating The Implications of Cyberattacks Against Precision Agricultural Equipment**

Mark Freyhof[1], George Grispos[2], Santosh K. Pitla[1], William Mahoney[2]
[1]University of Nebraska-Lincoln, USA
[2]University of Nebraska-Omaha, USA
mtf92015@gmail.com
ggrispos@unomaha.edu
spitla2@unl.edu
wmahoney@unomaha.edu

**Abstract[1]:** As various technologies are integrated and implemented into the food and agricultural industry, it is increasingly important for stakeholders throughout the sector to identify and reduce cybersecurity vulnerabilities and risks associated with these technologies. However, numerous industry and government reports suggest that many farmers and agricultural equipment manufacturers do not fully understand the cyber threats posed by modern agricultural technologies, including CAN bus-driven farming equipment. This paper addresses this knowledge gap by attempting to quantify the cybersecurity risks associated with cyberattacks on farming equipment that utilize CAN bus technology. The contribution of this paper is twofold. First, it presents a hypothetical case study, using real-world data, to illustrate the specific and wider impacts of a cyberattack on a CAN bus-driven fertilizer applicator employed in row-crop farming. Second, it establishes a foundation for future research on quantifying cybersecurity risks related to agricultural machinery.

## 1. Introduction

An emerging technological trend in the food and agricultural industry is the implementation of Controller Area Network (CAN) bus networks and telematics within farming equipment (Freyhof et al., 2022). This technology helps a farmer to diagnose equipment problems and to monitor the field performance of a machine (Pitla et al., 2014). However, these CAN bus networks also extend the attack surfaces of this equipment and introduce cybersecurity challenges into an industry already being targeted by cybercriminals. For example, a cyberattack targeting beef processing plants in the United States (U.S.) caused a fifth of these plants to shut down production and resulted in one organization paying an $11 million ransom (Bunge, 2021). Further complicating matters, the Federal Bureau of Investigation (2022) suggests that cybercriminals are more likely to attack farmers during two critical periods: *during the planting and harvesting seasons*, which can result in negative impacts on the food supply chain.

Previous research (Wright, 2011) has demonstrated that the CAN bus network within on-road vehicles can be exploited, allowing an attacker to gain control of a vehicle remotely. This same attack is likely possible in farm equipment that uses CAN bus networks; an attacker might attempt to exploit the CAN bus and change the application rates of seeds, fertilizer, and chemicals. If a tractor is connected to an implement (e.g., a fertilizer applicator) through the CAN bus, an attacker can potentially compromise and attack the connected implement. According to the Department of Homeland Security (2018), this type of attack is plausible, and would likely result in crop damage and yield losses.

There are concerns that farmers and agriculture technology manufacturers do not fully understand the cyber threats introduced by precision agriculture and that further research is needed to investigate and understand the impact of attacks targeting farm equipment (Department of Homeland Security, 2018). This research is a first step. The contribution of this paper is twofold. First, the research presents a hypothetical case study, using real-world crop production scenarios, to demonstrate the impacts of cyberattacks on farm equipment. Second, it presents a foundation for future research in this domain. The rest of the paper is structured as follows: Section 2 presents previous work related to cybersecurity in farming and agricultural settings. Section 3 presents case study background information. Section 4 presents the research methodology. Section 5 presents the findings and section 6 discusses the implications of the findings in the context of farm settings. Section 7 concludes the research and presents ideas for future work.

---

[1] Please cite this pre-print as: M. Freyhof, G. Grispos, S.K. Pitla, W. Mahoney (2025). Investigating The Implications of Cyberattacks Against Precision Agricultural Equipment, 20th International Conference on Cyber Warfare and Security (ICCWS 2025), Williamsburg, Virginia, USA

## 2. Cybersecurity Challenges in Farming and Agriculture

While farmers have embraced various technologies, an analysis of the literature suggests that the integration of these technologies introduces cybersecurity concerns. Jahn et al. (2019) argue that increased use of smart farm equipment will increase the risk of cyber incidents, impacting the planting, cultivating, and harvesting of farm products. Sontowski et al. (2020) discuss cyberattacks involving computer networks within farms, specifically the 802.11 wireless protocol, and demonstrate how an attacker could disrupt sensor data stored in the cloud. Alahmadi et al. (2022), and Gupta et al. (2020), add that Internet of Things devices integrated into farms could also be compromised for malicious benefit.

To address some of these concerns, Chi, et al. (2017) developed a smart farming cybersecurity framework that includes detecting sensor events, access control, and encryption to prevent cyberattacks. Barreto and Amaral (2018) argue that farm settings require a multi-disciplinary effort including policy definition and education. Researchers have also developed several cybersecurity testbeds to help develop countermeasures for farming settings. Freyhof (2022) proposed a testbed for investigating wireless vulnerabilities in farming robots, while Agarwal, et al. (2022) developed a testbed specifically for dairy farms.

Studies also suggest that farmers are not aware of the risks associated with implementing new technologies. Nikander et al. (2020) studied computer networks at six Finnish dairy farms and identified several issues, including the use of home equipment, limited firewalls, and default configurations on devices. Spaulding and Wolf (2018) investigated cybersecurity awareness and training demands among farmers in Illinois. According to these findings, farmers with ten or fewer years of farming experience were more likely to use technology, but only 10% of the surveyed individuals indicated that they had attended any formal cybersecurity training. Spaulding and Wolf concluded that many farmers lack the education they require to identify and mitigate the growing number of cyber threats and risks that impact the farming industry. To help address this educational concern, Grispos, et al. (2025) propose a cybersecurity curriculum that aims to create a more secure workforce in the food and agricultural sector. While previous research has highlighted high-level cybersecurity threats in farming and agricultural settings, minimal research has investigated the financial cost and impact of security incidents and cybercrime affecting farming equipment implemented in today's modern farm.

## 3. Background

Nearly 24% of the U.S. cash crop industry is commodity corn production; U.S. farmers grew 13.7 billion corn bushels in 2022 (United States Department of Agriculture, 2022, United States Department of Agriculture, 2023). To produce such yields, farmers must obtain a delicate balance between the correct amount of water, soil, light, and nutrients in the production process. Research has also shown a correlation between available nitrogen fertilizer and higher corn crop yields (Shapiro et al., 2019). However, proper management of this nutrient is often required to prevent its loss, which can result in negative environmental impacts and profit financial losses. Another complicating factor is the rising costs associated with nitrogen fertilizer, which has nearly doubled in recent years (Quinn, 2023).

Most corn producers rely on synthetic fertilizers (Sellars and Nunes, 2021), applied at *planting time* and *in-season*; with a farmer utilizing multiple methods to achieve this goal, including broadcast spreaders, injection equipment, and side dress equipment (Grift and Hofstee, 2002, Dellinger et al., 2008, Laguë, 1991, Blaylock et al., 2005). This research focuses on *side dress equipment*, which spreads fertilizer between the field rows. This equipment has evolved into electrically controlled units via CAN bus networks, allowing a farmer to send commands from a tractor (Case IH, 2019), which allows for controlling the fertilizer pressure and flow rate. Flow meter gauges can then provide feedback on the amount applied, usually displayed on a screen in the tractor cabin. The importance of fertilizer when growing corn, known CAN bus vulnerabilities, and the increasing costs related to purchasing fertilizer make this a prime setting for potential exploitation.

## 4. Research Methodology

Based on the above, it is hypothesized that *a cyberattack on a CAN-bus side dress fertilizer equipment will negatively impact profits associated with corn crops that are treated using the equipment.* Hence, this research investigates a hypothetical financial impact that results when side-dress equipment is successfully attacked. This type of attack has previously been discussed in a Department of Homeland Security report (2018).

A case study is developed using a hypothetical corn producer ("the farmer") in Saunders County, Nebraska. This location is selected due to historically successful corn yields (United States Department of Agriculture, 2021). The farmer uses CAN bus-driven fertilizer side-dress equipment on a 100-acre corn field (hereafter referred to as "the field") with 1-acre zones. For convenience, refer to this 10x10-acre area, with A..J for columns and 1..10 for rows, giving a 100-acre grid.

To investigate the hypothesis, the case study uses equations (see Appendix) from the Nebraska Nitrogen Formula (Shapiro et al., 2019). The formulas in Equations 1 and 2 allow one to set an expected yield, as well as fertilizer application timing and price adjustment. Data used in the case study is drawn from public information related to corn farmers in Saunders County. Yield goals, soil nitrate levels, and soil organic matter levels are also based on information from the United States Department of Agriculture and the State of Nebraska (Wortmann et al., 2017). While corn prices vary, the price of corn is set at $7.53, and the price of fertilizer at $1.10/lb, which is based on prices at the time of the study. Additional data used in the case study is described in the control case and cyberattack descriptions.

### 4.1. Control Case

To compare to a cyberattack, first consider a *normal* in-season fertilizer application operation, without the cyberattack. Yield goals are randomly allocated between 150-190 bushels/acre, based on historic values in Saunders County (United States Department of Agriculture, 2018). The yield values for each zone are shown in Figure 1a. The field *should* produce 16,983 bushels. In this control case, a 30-bushel yield boost above expected yield values will be the maximum attainable yield, if fertilizer is overapplied above the prescribed rates. The minimum attainable yield value will be 100 bushels if no fertilizer is applied.

Next, allocate soil nitrate levels for each 1-acre zone, using known values for central Nebraska (Shapiro et al., 2019). Each zone has a value between 2.0ppm – 4.0ppm (1b) and a soil organic matter value (Figure 1c), between 1.8% and 2.2%. The timing factor for the field is *0.95* since fertilizer applications will be both at-planting and in-season. Based on the prices above, the *price adjustment factor* is *0.926*.

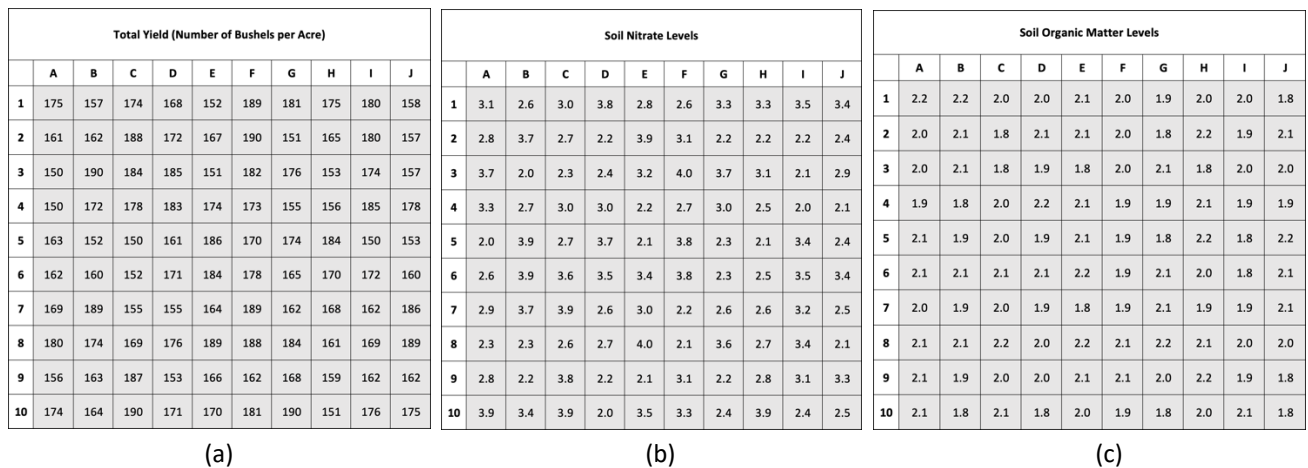

**Figure 1:** Formula input values for the hypothetical field

Now utilize Eq.1 to calculate the fertilizer needs ($N_{rec}$) per zone, e.g. for zone A1, $N_{rec}$ = 147 lb./acre. The approach is used to calculate total fertilizer recommendations for each acre. Using recommendations (Shapiro et al., 2019), the applications for the field are split, *25% at planting* and *75% during in-season* (Figures 2a and 2b, respectively). The entire field requires 3,690 lbs. of fertilizer during planting and 11,069 lbs. of fertilizer during the in-season period.

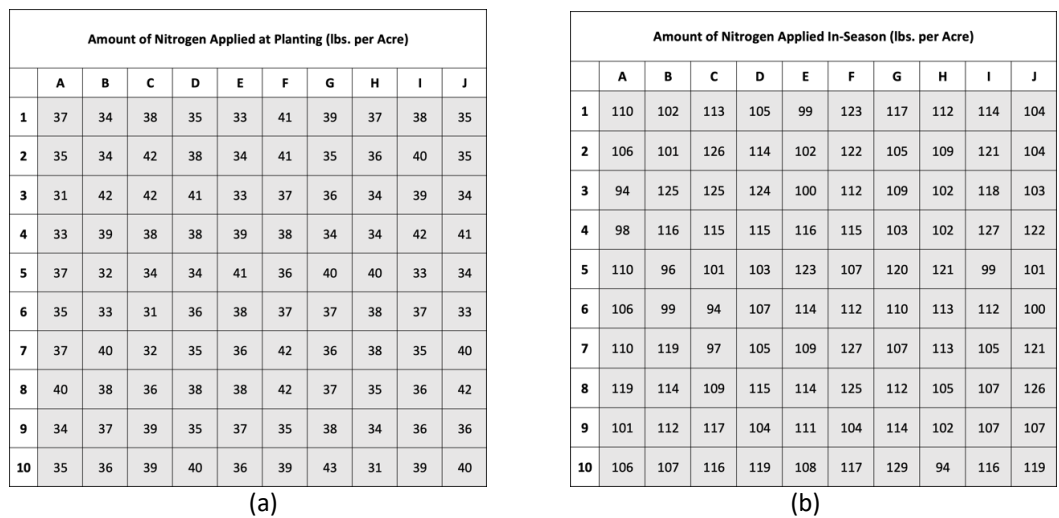

**Figure 2:** Nitrogen input (lbs./acre) for control scenario (a) at planting and (b) in-season

Finally, calculate the field's potential profit loss/gain ($P_{LG}$), the difference between the Expected Yield ($EY$) and the Actual Yield ($AY$) using Eq.3 and Eq.4. Since this is the control case, $AY = EY$, 16,983 bushels. Use Eq.5-11 to calculate potential profit loss or gain ($P_{LG}$). The total *expected* revenue for the control case is $127,881.99, and the cost of fertilizer is $16,234.90. Eq.7 gives a total profit of $111,647.09. As this is the control case, the total actual profit is the same, as the farmer has not yet been impacted by a cyberattack.

### 4.2 Cyberattack Descriptions

The cyberattack *modus operandi* is to target a tractor and side-dress fertilizer equipment (hereafter referred to as "the implement") connected via a CAN bus. The farmer programs the implement to apply the correct amount of fertilizer during initial planting, but the attack will cause financial losses by increasing/decreasing the fertilizer applied during the *in-season application.* Three attack scenarios are proposed.

**Scenario 1**: The attacker modifies the amount of fertilizer applied to specific zones, either *decreasing* by 25% or 50%, *leaving at the original amount*, or *increasing* by 50% or 100%. For example, some zones will receive more fertilizer, some less, and some zones receive the correct amount. Figure 3a shows the percentage of the fertilizer applied to a zone; for example, B5 receives 50% of the prescribed rate.

**Scenario 2**: The attacker attempts to reduce the amount of fertilizer in zones A1 to D10 and F1 to I10, applying only 45% of the required fertilizer. They then apply 280% of the needed fertilizer in rows E and J. Fertilizer is wasted in two rows and a reduced amount is applied in the remaining rows (Figure 3b).

**Scenario 3**: In Figure 3c, this attack involves adjusting the fertilizer in certain zones – 25% of the prescribed rate for zones B1-10, D1-10, F1-10, and H1-10, 100% of the prescribed rate for zones A1-10, E1-10, and K1-10, and 200% of the prescribed rate for zones C1-10, G1-10, and I1-10.

For each scenario, we consider:

- **The amount of fertilizer used and the cost of applying this fertilizer, after the cyberattack**. This is calculated by comparing the expected fertilizer applied versus the actual amount due to the cyberattack.

- **The actual corn yield, after the cyberattack**. This is calculated by comparing the yield produced by the field after the cyberattack, with the expected yield without a cyberattack.

- **The financial impact, after the cyberattack.** This is calculated by comparing the expected profit from the control case with the profit generated by the corn produced after the cyberattack.

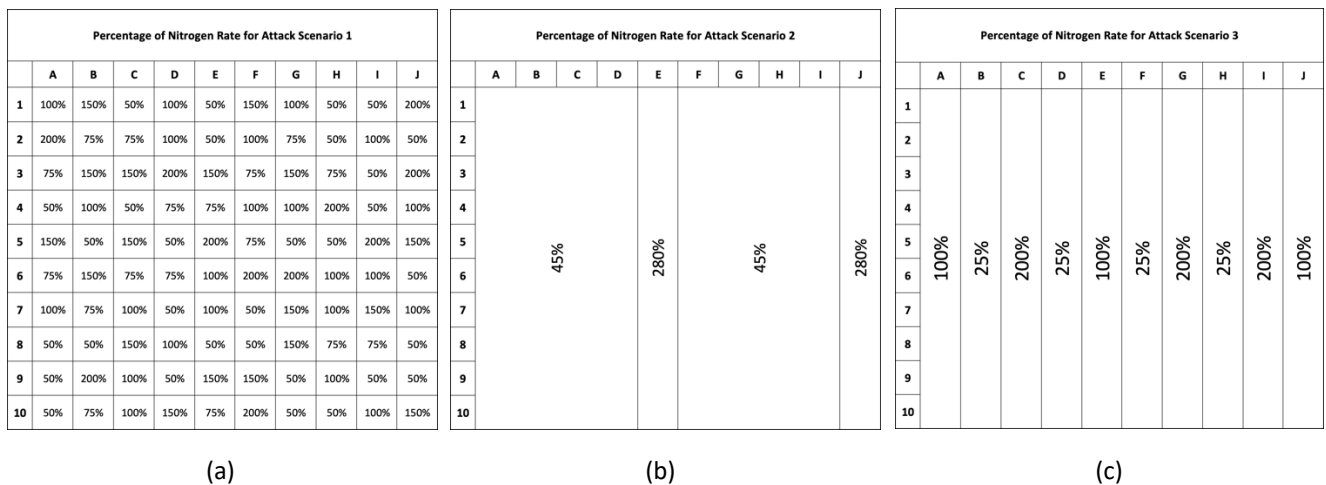

**Figure 3:** Attack scenarios

### 4.3 Limitations

Assume first that the farmer uses a variable-rate nitrogen side-dress implement, connected to a tractor via CAN bus. Assume that the tractor can be compromised using approaches discussed in the literature (Wright, 2011, Miller and Valasek, 2015). The implement can increase fertilizer application pressure and flow rates and support application outlets that allow for a large variation in application rates. The study is again based on real-world values for Saunders County, Nebraska, fertilizer cost, and the price of corn at the time of the research.

## 5. Findings

All three attack scenarios result in a deviation in the cost of fertilizer applied to the field, and thus differences in the financial profits obtained from the corn yields after each attack. The following subsections discuss the specific impact of the attack scenarios on the hypothetical field.

### 5.1 Scenario 1

Here, 11,093 lbs. of fertilizer are used in-season. The total amount applied after this attack is 24 lbs. more fertilizer than the control scenario. Figure 4 shows the amount of fertilizer applied after the attack (yellow is the same, red is more, green is less). Using Eq.4, we can calculate the number of bushels produced, presented in Figure 5 (same colours), as compared to the control case. The yield after this attack is 15,243 bushels, 1,740 bushels *fewer* than the expected yield. This is a 17.4-bushel yield penalty per acre, and the total revenue after the attack is $114,779.79.

**Amount of Nitrogen Applied In-Season After Attack Scenario 1 (lbs. per Acre)**

|    | A   | B   | C   | D   | E   | F   | G   | H   | I   | J   |
|----|-----|-----|-----|-----|-----|-----|-----|-----|-----|-----|
| 1  | 110 | 153 | 57  | 105 | 50  | 185 | 117 | 56  | 57  | 208 |
| 2  | 213 | 76  | 95  | 114 | 51  | 122 | 79  | 54  | 121 | 52  |
| 3  | 71  | 188 | 188 | 248 | 150 | 84  | 163 | 77  | 59  | 206 |
| 4  | 49  | 116 | 58  | 86  | 87  | 115 | 103 | 205 | 63  | 122 |
| 5  | 166 | 48  | 151 | 51  | 247 | 80  | 60  | 60  | 197 | 152 |
| 6  | 79  | 148 | 71  | 80  | 114 | 224 | 220 | 113 | 112 | 50  |
| 7  | 110 | 89  | 97  | 52  | 109 | 64  | 160 | 113 | 158 | 121 |
| 8  | 59  | 57  | 164 | 115 | 57  | 63  | 168 | 79  | 81  | 63  |
| 9  | 51  | 223 | 117 | 52  | 166 | 155 | 57  | 102 | 54  | 53  |
| 10 | 53  | 80  | 116 | 179 | 81  | 234 | 64  | 47  | 116 | 179 |

**Number of Bushels per Acre After Attack Scenario 1**

|    | A   | B   | C   | D   | E   | F   | G   | H   | I   | J   |
|----|-----|-----|-----|-----|-----|-----|-----|-----|-----|-----|
| 1  | 175 | 187 | 104 | 168 | 100 | 219 | 181 | 106 | 110 | 188 |
| 2  | 191 | 131 | 150 | 172 | 103 | 190 | 119 | 100 | 180 | 100 |
| 3  | 121 | 220 | 214 | 215 | 181 | 147 | 206 | 122 | 102 | 187 |
| 4  | 100 | 172 | 107 | 146 | 138 | 173 | 155 | 186 | 108 | 178 |
| 5  | 193 | 100 | 180 | 100 | 216 | 137 | 102 | 108 | 180 | 183 |
| 6  | 129 | 190 | 122 | 137 | 184 | 208 | 195 | 170 | 172 | 100 |
| 7  | 169 | 152 | 155 | 100 | 164 | 111 | 192 | 168 | 192 | 186 |
| 8  | 106 | 102 | 199 | 176 | 117 | 110 | 214 | 128 | 136 | 111 |
| 9  | 100 | 193 | 187 | 100 | 196 | 192 | 100 | 159 | 100 | 100 |
| 10 | 107 | 132 | 190 | 201 | 137 | 211 | 112 | 100 | 176 | 205 |

**Figure 4:** Amount of fertilizer after attack 1  **Figure 5:** Yield per acre after attack 1

The farmer would lose $13,102.20 in corn sales across the field here, $131 per acre. Since 24 additional pounds of fertilizer is used, the farmer pays $26.40 more than the control scenario. The total profit generated by the farmer after Scenario 1 is $98,518.49, $13,129 less than the control scenario.

### 5.2 Scenario 2

In this attack, the total amount of fertilizer applied is 14,734 lbs., with 11,044 lbs. used during the in-season application, 25 lbs. less than the control scenario. Figure 6 shows the fertilizer (red is more, green less) applied to each zone after Scenario 2. Figure 7 shows the achieved corn yield per zone (same colours) after the attack, compared to the control case. The yield of Scenario 2 is 4,291 bushels *fewer* than the expected yield, a 42.9 penalty per acre. Based on a yield total of 12,692 bushels, the total revenue for the field is $95,570.76.

| Amount of Nitrogen Applied In-Season After Attack Scenario 2 (lbs. per Acre) | | | | | | | | | | |
|---|---|---|---|---|---|---|---|---|---|---|
|  | A | B | C | D | E | F | G | H | I | J |
| 1 | 61 | 56 | 62 | 58 | 277 | 68 | 65 | 62 | 63 | 291 |
| 2 | 59 | 56 | 69 | 63 | 286 | 67 | 58 | 60 | 67 | 291 |
| 3 | 52 | 69 | 69 | 68 | 281 | 62 | 60 | 56 | 65 | 288 |
| 4 | 54 | 64 | 63 | 63 | 323 | 63 | 57 | 56 | 70 | 341 |
| 5 | 61 | 53 | 55 | 57 | 345 | 59 | 66 | 66 | 54 | 283 |
| 6 | 58 | 54 | 52 | 59 | 320 | 61 | 60 | 62 | 62 | 281 |
| 7 | 60 | 66 | 53 | 58 | 306 | 70 | 59 | 62 | 58 | 340 |
| 8 | 65 | 63 | 60 | 63 | 318 | 69 | 62 | 58 | 59 | 353 |
| 9 | 56 | 61 | 65 | 57 | 310 | 57 | 63 | 56 | 59 | 298 |
| 10 | 59 | 59 | 64 | 66 | 301 | 64 | 71 | 52 | 64 | 334 |

| Number of Bushels per Acre After Attack Scenario 2 | | | | | | | | | | |
|---|---|---|---|---|---|---|---|---|---|---|
|  | A | B | C | D | E | F | G | H | I | J |
| 1 | 112 | 100 | 111 | 110 | 182 | 120 | 117 | 113 | 117 | 188 |
| 2 | 102 | 105 | 120 | 107 | 197 | 122 | 100 | 103 | 113 | 187 |
| 3 | 100 | 119 | 116 | 117 | 181 | 119 | 115 | 100 | 109 | 187 |
| 4 | 100 | 109 | 114 | 117 | 204 | 110 | 100 | 100 | 116 | 208 |
| 5 | 101 | 100 | 100 | 105 | 216 | 111 | 109 | 115 | 100 | 183 |
| 6 | 102 | 105 | 100 | 111 | 214 | 116 | 103 | 107 | 111 | 190 |
| 7 | 108 | 123 | 101 | 100 | 194 | 119 | 102 | 106 | 104 | 216 |
| 8 | 113 | 109 | 106 | 112 | 219 | 118 | 120 | 102 | 109 | 219 |
| 9 | 100 | 102 | 122 | 100 | 196 | 103 | 105 | 101 | 104 | 192 |
| 10 | 114 | 106 | 125 | 107 | 200 | 117 | 120 | 100 | 111 | 205 |

**Figure 6:** Amount of fertilizer after attack 2      **Figure 7:** Yield per acre after attack 2

In this attack scenario, the farmer expects to lose $32,311.23 in corn sales across the field, compared to the control scenario, a $323 loss per acre. Since the farmer uses 25 lbs. *less* fertilizer than the control, the farmer pays $16,207.40, saving $27.50. However, from a total profit viewpoint, the farmer only generates $79,363.36, or $32,284 less than the control scenario.

### 5.3 Scenario 3

For Scenario 3, the total amount of fertilizer applied after this attack is 14,796 lbs., with 11,106 lbs. used in-season, which equates to 37 additional pounds of fertilizer. Figure 8 shows the field zones where more (red), less (green), or the same (yellow) amount was applied due to the attack, and Figure 9 shows the achieved corn yield for each zone after the attack. The corn yield after Scenario 3 is 15,061 bushels, 1,922 bushels less than the expected yield. With a yield of 15,061 bushels, the total revenue is $113,409.33.

| Amount of Nitrogen Applied In-Season After Attack Scenario 3 (lbs. per Acre) | | | | | | | | | | |
|---|---|---|---|---|---|---|---|---|---|---|
|  | A | B | C | D | E | F | G | H | I | J |
| 1 | 110 | 26 | 226 | 26 | 99 | 31 | 235 | 28 | 228 | 104 |
| 2 | 106 | 25 | 252 | 28 | 102 | 31 | 211 | 27 | 243 | 104 |
| 3 | 94 | 31 | 251 | 31 | 100 | 28 | 218 | 26 | 236 | 103 |
| 4 | 98 | 29 | 230 | 29 | 116 | 29 | 206 | 26 | 253 | 122 |
| 5 | 110 | 24 | 201 | 26 | 123 | 27 | 239 | 30 | 197 | 101 |
| 6 | 106 | 25 | 189 | 27 | 114 | 28 | 220 | 28 | 224 | 100 |
| 7 | 110 | 30 | 194 | 26 | 109 | 32 | 214 | 28 | 210 | 121 |
| 8 | 119 | 29 | 219 | 29 | 114 | 31 | 225 | 26 | 215 | 126 |
| 9 | 101 | 28 | 235 | 26 | 111 | 26 | 228 | 26 | 214 | 107 |
| 10 | 106 | 27 | 232 | 30 | 108 | 29 | 257 | 23 | 232 | 119 |

| Number of Bushels per Acre After Attack Scenario 3 | | | | | | | | | | |
|---|---|---|---|---|---|---|---|---|---|---|
|  | A | B | C | D | E | F | G | H | I | J |
| 1 | 175 | 100 | 204 | 100 | 152 | 100 | 211 | 100 | 210 | 158 |
| 2 | 161 | 100 | 218 | 100 | 167 | 100 | 181 | 100 | 210 | 157 |
| 3 | 150 | 100 | 214 | 100 | 151 | 100 | 206 | 100 | 204 | 157 |
| 4 | 150 | 100 | 208 | 100 | 174 | 100 | 185 | 100 | 215 | 178 |
| 5 | 163 | 100 | 180 | 100 | 186 | 100 | 204 | 100 | 180 | 153 |
| 6 | 162 | 100 | 182 | 100 | 184 | 100 | 195 | 100 | 202 | 160 |
| 7 | 169 | 100 | 185 | 100 | 164 | 100 | 192 | 100 | 192 | 186 |
| 8 | 180 | 100 | 199 | 100 | 189 | 100 | 214 | 100 | 199 | 189 |
| 9 | 156 | 100 | 217 | 100 | 166 | 100 | 198 | 100 | 192 | 162 |
| 10 | 174 | 100 | 220 | 100 | 170 | 100 | 220 | 100 | 206 | 175 |

**Figure 8**: Amount of fertilizer after attack 3      **Figure 9:** Yield per acre after attack 3

The farmer loses $14,476.66 in corn sales. This scenario uses 37 lbs *more* fertilizer; the farmer pays $16,275.60, *$40.70 more*. The total profit generated by the farmer is $97,133.73, $14,513 less than the control scenario.

## 5.4 Results Summary and Discussion

Figure 10 summarizes the financial impact of the three attacks on the farmer's field, which shows the farmer sustaining financial losses in all three cases. In Scenario 1, the farmer would lose $13,128.60, in Scenario 2 the loss would be $32,283.73 and in Scenario 3 the farmer would lose $14,513.36. In two out of the three scenarios (1 and 3), *more* fertilizer would be applied, increasing the costs. While *less* fertilizer would be applied in Scenario 2, the decrease in the amount of fertilizer results in a larger overall financial loss. If more fertilizer is used during the attack, a farmer may notice; hence, it is reasonable to hypothesize that the "best" cyberattack is to under-fertilize.

The FBI has raised concerns that attackers are increasingly likely to target farmers during planting and harvesting (Federal Bureau of Investigation, 2022). These findings reinforce these concerns. As further technologies are integrated into farming practices, crop yield targets, soil measurements, and recommended fertilizer rates will continue to be more site-specific. This provides an opportunity for attackers to exploit the technologies driving these practices and cause financial damage. As shown in the findings from this research, any disruptions to fertilizer input caused by a cyberattack could result in monetary losses.

Since farmers are becoming increasingly reliant on digital monitoring technologies (Freyhof, 2022), any changes in fertilizer application rates may go undetected, as indicators may show normal operations. If the cybercriminal also takes control of the display unit (either in the cabin or on the implement itself), the first indication of the cyberattack might only emerge when the crop is finally harvested, and a low crop yield is identified. Another concern is that it is entirely likely that many farmers might use the same type of equipment, potentially putting an entire country/area at risk if this implement is susceptible to this type of attack or if a vulnerability is identified and exploited in the implement itself.

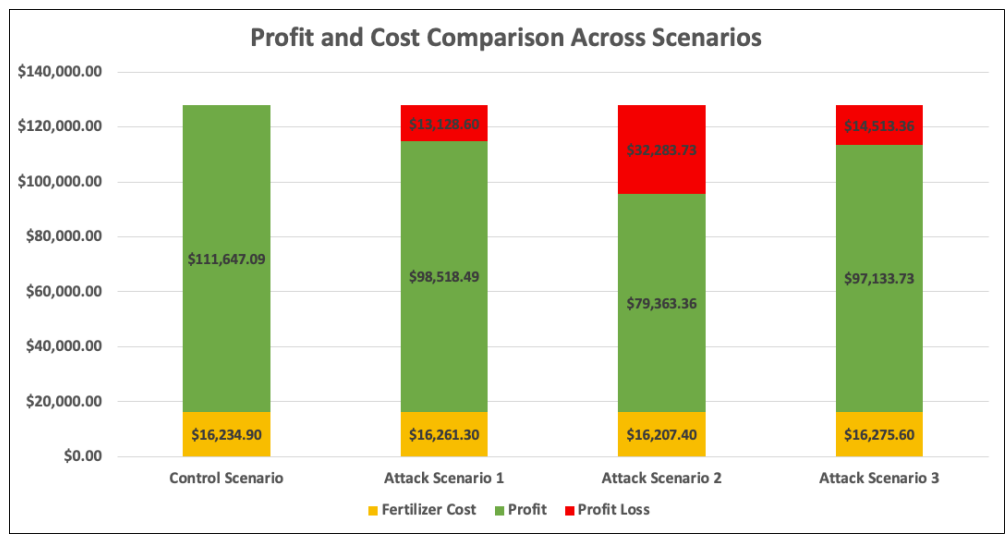

**Figure 10:** Results summary

Cyberattacks against technology-driven farming equipment prompts the discussion that security incident investigations of this equipment may soon be required to establish root causes and identify controls for prevention (Grispos et al., 2014). Some standards (e.g., NIST 800-61 Computer Security Incident Handling Guide (Cichonski et al., 2012)) have been developed for investigating traditional computer systems, but do not consider the complexity of farm systems. For example, NIST 800-61 states that when incidents occur, impacted systems should be isolated, powered down, and data collected. But if farming equipment is powered down, what data remains? If any data does remain, does it help investigators, and how can it be collected? Tools and techniques recommended for collecting and analysing data are likely invalidated in farming and agricultural contexts. For example, if incident handlers use tools such as 'FTK Imager' to create copies of storage devices, would these tools work with farm equipment? If not, alternative techniques that could include examining potential memory chips (e.g. (Grispos et al., 2021) need to be created. Lastly, the discussion surrounding security incident response in farming and agricultural settings raises the question: who handles the incident? If farmers rely on renting equipment from third parties (Brennan et al., 2008), is it the manufacturer, the farmer, or the third party renting out equipment?

# 6. Conclusions And Future Research

As further technologies are integrated into farming and agricultural contexts, it becomes important for farmers to understand the vulnerabilities and risks associated with these technologies. This research investigates the risks associated with three hypothetical cyberattacks against CAN bus-driven equipment used to disperse fertilizer in a field. From a financial perspective, a farmer would sustain financial losses in all three attacks. There are several potential avenues for future research. Further research is needed to investigate and understand the actual risks that could be exploited by attackers, along with developing models of the damage these attacks could cause to farmers and food production. The importance of the food industry to local, regional, and state economies should not be underestimated. Cybersecurity risk frameworks and best practices will need to be developed for the farming and agricultural technology ecosystem.

This study can be repeated and extended to consider other farming scenarios, equipment, and geographical locations. While this research focused on cyberattacks against fertilizer equipment used in Nebraska, the study could be repeated in evaluating the cybersecurity risks of Internet-connected tractors, Internet of Things devices, drones, farming applications and services, and so on. This extended study will investigate the financial impact of cyberattacks against these technologies. The findings from the extended study could help determine the cost and financial investment needed to improve cybersecurity within agriculture. There is also a need to develop cybersecurity countermeasures and controls to help enhance the security of farming equipment. These countermeasures can focus on the technologies used in farms and agricultural settings, as well as the individuals who work in these environments. Finally, future work could develop and evaluate cybersecurity best practices and standards for agriculture, which would help ensure that a baseline level of cybersecurity exists for many economies around the world.

# Appendix

$$N_{rec} = [35 + (1.2 \times EY) - (8 \times NO_3 - N\ ppm) - (0.14 \times EY \times OM) - other\ N\ credits] \times Price_{adj} \times Timing_{adj} \quad \text{(Eq.1)}$$

Where:
$N_{rec}$ = recommended nitrogen input for corn (lb./ac)
$EY$ = expected yield (bu/ac)
$NO_3 - N\ ppm$ = average nitrate-N concentration, root zone, parts per million
$OM$ = percent soil organic matter (0.5%-3%)
$other\ N\ credits$ = include N from prior legume crop, manure and organic material applied, and irrigation water N.
$Price_{adj}$ = price adjustment coefficient
$Timing_{adj}$ = adjustment factor for fall, spring, and split applications = 0.95 for split application

$$Price_{adj} = 0.263 + \left(0.1256 * \frac{P_{corn}}{P_{nitrogen}}\right) - \left(0.00421 * \left(\frac{P_{corn}}{P_{nitrogen}}\right)^2\right) \quad \text{(Eq.2)}$$

Where:
$Price_{adj}$ = price adjustment coefficient
$P_{corn}$ = price of corn ($/bu)
$P_{nitrogen}$ = price of nitrogen ($/lb.)

To calculate the resulting yield, the Nebraska Nitrogen Formula needs to be rearranged to solve for Expected ($EY$) Yield and Actual Yield ($AY$), as shown in Eq.3 and Eq.4.

$$\boldsymbol{EY} = \max\left[EY_{min},\ \min\left[EY_{max}, \frac{\left(\frac{N_{rate}}{Price_{adj} \times Timing_{adj}} + (8*NO_3 - N\ ppm) - 35 + other\ N\ credits\right)}{1.2 - (0.14*OM)}\right]\right] \quad \text{(Eq.3)}$$

Where:
$EY$ = expected yield (bu/ac)
$EY_{min}/EY_{max}$ = minimum/maximum expected yield (bu/ac)
$N_{rate}$ = recommended nitrogen for corn (lb./ac)
$NO_3 - N\ ppm$ = average nitrate-N concentration in the root zone in parts per million
$OM$ = percent soil organic matter (0.5%-3%)
$other\ N\ credits$ = include N from prior legume crop, manure and organic material applied, and irrigation water N.
$Price_{adj}$ = price adjustment coefficient
$Timing_{adj}$ = adjustment factor for fall, spring, and split applications = 0.95 for split application

$$AY = \max\left[EY_{min},\ \min\left[EY_{max},\ \frac{\left(\frac{N_{rate,act}}{Price_{adj} \times Timing_{adj}} + (8*NO_3-N\ ppm) - 35 + other\ N\ credits\right)}{1.2 - (0.14*OM)}\right]\right]$$ (Eq.4)

Where:
$AY$ = actual yield (bu/ac)
$EY_{min}$ = minimum expected yield (bu/ac)
$EY_{max}$ = maximum expected yield (bu/ac)
$N_{rate,act}$ = actual nitrogen input for corn grain (lb./ac)
$NO_3 - N\ ppm$ = average nitrate-N concentration in the root zone in parts per million
$OM$ = percent soil organic matter (0.5%-3%)
$other\ N\ credits$ = include N from previous legume crop, manure and other organic material applied, and irrigation water N.
$Price_{adj}$ = price adjustment coefficient
$Timing_{adj}$ = adjustment factor for fall, spring, and split applications = 0.95 for split application

$$R_{exp} = EY_{tot} * v_{corn}$$ (Eq.5)

Where:
$R_{exp}$ = expected revenue ($)
$EY_{tot}$ = expected total yield (bu)
$v_{corn}$ = value of corn grain ($/bu)

$$C_{exp} = N_{tot,exp} * c_{nitrogen}$$ (Eq.6)

Where:
$C_{exp}$ = expected total cost ($)
$N_{tot,exp}$ = expected total pounds of nitrogen to be applied (lb.)
$c_{nitrogen}$ = cost of nitrogen per pound ($/lb.)

$$P_{exp} = R_{exp} - C_{exp}$$ (Eq.7)

Where:
$P_{exp}$ = total expected profit ($)
$R_{exp}$ = expected revenue ($)
$C_{exp}$ = expected total cost ($)

$$R_{act} = AY_{tot} * v_{corn}$$ (Eq.8)

Where:
$R_{act}$ = actual revenue ($)
$AY_{tot}$ = actual total yield (bu)
$v_{corn}$ = value of corn grain ($/bu)

$$C_{act} = N_{tot,act} * c_{nitrogen}$$ (Eq.9)

Where:
$C_{act}$ = actual total cost ($)
$N_{tot,act}$ = actual total pounds of nitrogen to be applied (lb.)
$c_{nitrogen}$ = cost of nitrogen per pound ($/lb.)

$$P_{act} = R_{act} - C_{act}$$ (Eq.10)

Where:
$P_{act}$ = total actual profit ($)
$R_{act}$ = actual revenue ($)
$C_{act}$ = actual total cost ($)

$$P_{LG} = P_{exp} - P_{act}$$ (Eq.11)

Where:
$P_{LG}$ = profit loss or gain ($)
$P_{exp}$ = total expected profit ($)
$P_{act}$ = total actual profit ($)

## Research Funding

This study was partially funded by the University of Nebraska Collaboration Initiative, under the grant "Security and Hackability Considerations of Driverless Tractors and Agricultural Robots".